%% file: manuscript.tex
\documentclass[letterpaper, conference, 10pt]{ieeeconf}
                \author{Pau de las Heras Molins\(^{*}\), Eric Roy-Almonacid\(^{*}\), Dong Ho Lee, Lasse Peters,\\ David Fridovich-Keil, and Georgios Bakirtzis \thanks{\(^{*}\text{Equal contribution and corresponding authors.}\)}\thanks{P.~de las Heras Molins (pau.de.las.heras@estudiantat.upc.edu) and E.~Roy-Almonacid (eric.roy@upc.edu) are with  Universitat Politècnica de Catalunya. D.~H.~Lee and D.~Fridovich-Keil are with The University of Texas~at~Austin. L.~Peters is with Technische Universiteit Delft. G. Bakirtzis is with~LTCI, Télécom~Paris, Institut Polytechnique de Paris.}}
\IEEEoverridecommandlockouts
\usepackage{etoolbox}
\AtBeginDocument{\frenchspacing}
\usepackage[final,nopatch=footnote]{microtype}
\usepackage{stix}
\usepackage{circledsteps}
\usepackage[noend,noline,plainruled]{algorithm2e}
\DontPrintSemicolon
\SetAlgoCaptionSeparator{.}
\SetAlCapNameFnt{\footnotesize}
\SetAlCapFnt{\mdseries \footnotesize}
\SetAlCapSkip{0.5em}
\SetAlgoInsideSkip{medskip}
\SetKwSty{textsc}
\SetKwInput{KwData}{\mdseries\textsc{data}}
\SetKwInput{KwResult}{\mdseries\textsc{result}}
\usepackage{hyperref}
\hypersetup{colorlinks=true, citecolor=gray, linkcolor=gray, urlcolor=gray}
\let\labelindent\relax
\usepackage{enumitem}
\input{glyphtounicode}
\usepackage{multirow,booktabs,rotating,graphicx,import}
\makeatletter
\let\MYcaption\@makecaption
\makeatother
\usepackage[font=footnotesize]{subcaption}
\makeatletter
\let\@makecaption\MYcaption
\makeatother

\BeforeBeginEnvironment{tabular*}{\footnotesize}
\AtBeginDocument{\allowdisplaybreaks[1]}
\AtBeginEnvironment{tabular}{}
\makeatletter\g@addto@macro\@floatboxreset\centering\makeatother
\usepackage{transparent}
\usepackage{amsmath,mathtools}

\usepackage{annotate-equations}
\tikzset{annotate equations/arrow/.style={-, shorten <=2pt}}
\DeclareMathOperator*{\minimize}{minimize}
\usepackage{siunitx}
\DeclarePairedDelimiter{\abs}{\lvert}{\rvert}
\usepackage[doi=true,isbn=false,url=true,eprint=false,style=ieee,citestyle=numeric-comp, datamodel=software,maxbibnames=99]{biblatex}
\DeclareDelimFormat{multicitedelim}{\addcomma\,}
\DeclareDelimFormat[bib]{multicitedelim}{\addcomma\,}

\usepackage{software-biblatex}
\preto{\cite}{\unskip~}
\makeatletter
\patchcmd{\@IEEEyesnumber}
{\stepcounter}
{\refstepcounter}
{}{}
\patchcmd{\@@IEEEeqnarray}
{\stepcounter}
{\refstepcounter}
{}{}
\patchcmd{\@@IEEEeqnarraycr}
{\stepcounter{IEEEsubequation}}
{\refstepcounter{IEEEsubequation}}
{}{}
\patchcmd{\@@IEEEeqnarraycr}
{\stepcounter{IEEEsubequation}}
{\refstepcounter{IEEEsubequation}}
{}{}
\patchcmd{\@@IEEEeqnarraycr}
{\stepcounter{IEEEequation}}
{\refstepcounter{IEEEequation}}
{}{}
\patchcmd{\@@IEEEeqnarraycr}
{\stepcounter{IEEEequation}}
{\refstepcounter{IEEEequation}}
{}{}
\makeatother
\usepackage[capitalise]{cleveref}
\usepackage{ifthen}
\makeatletter
\newcounter{IEEE@bibentries}
\renewcommand\IEEEtriggeratref[1]{%
\renewbibmacro{finentry}{%
\stepcounter{IEEE@bibentries}%
\ifthenelse{\equal{\value{IEEE@bibentries}}{#1}}
{\finentry\@IEEEtriggercmd}
{\finentry}%
}%
}
\makeatother
\usepackage{balance}
\newcommand{\argmin}{\mathop{\textrm{argmin}}}

\usepackage{layouts}
\usepackage{pgfplots}
\usepackage{amsmath}
\pgfplotsset{compat=newest}
\usepackage{roboto}
\RequirePackage[l2tabu,orthodox]{nag}
\PassOptionsToPackage{x11names}{xcolor}
\date{\today}
\title{\LARGE \textbf{Approximate solutions to games of ordered preference}}
\hypersetup{
 pdfauthor={},
 pdftitle={Approximate solutions to games of ordered preference},
 pdfkeywords={},
 pdfsubject={},
 pdfcreator={},
 pdflang={English}}
\begin{document}

\maketitle
\begin{abstract}
Autonomous vehicles must balance ranked objectives, such as minimizing travel time, ensuring safety, and coordinating with traffic. Games of ordered preference effectively model these interactions but become computationally intractable as the time horizon, number of players, or number of preference levels increase. While receding horizon frameworks mitigate long-horizon intractability by solving sequential shorter games, often warm-started, they do not resolve the complexity growth inherent in existing methods for solving games of ordered preference. This paper introduces a solution strategy that avoids excessive complexity growth by approximating solutions using lexicographic iterated best response (IBR) in receding horizon, termed ``lexicographic IBR over time.'' Lexicographic IBR over time uses past information to accelerate convergence. We demonstrate through simulated traffic scenarios that lexicographic IBR over time efficiently computes approximate-optimal solutions for receding horizon games of ordered preference, converging towards generalized Nash equilibria.
\end{abstract}
\section{Introduction}
\label{sec:org4fb4ae8}

Complex agent decisions are often characterized by conflicting objectives.
An autonomous vehicle, for example, must avoid collisions while also staying
on the road, reaching a goal position,
and obeying the speed limit.
This problem is even
harder when objectives of multiple agents conflict.
Agents must determine which strategies are preferable,
for example, going off-road
to prevent a crash with another vehicle.
Such decisions reflect an underlying comparative evaluation
that ranks possible trajectories according
to subjective preferences.
We study the problem
of comparative evaluations
in transportation systems 
from the lens of \emph{preference relations}
that codify prioritized metrics; metrics
that human drivers implicitly follow on the road.

\begin{figure}[!t]
\def\svgwidth{\linewidth}
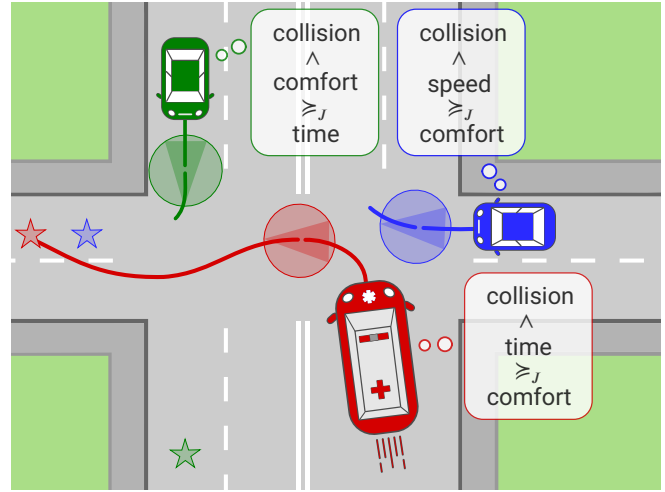
\caption{Agents operate under shared constraints, but are incentivized by different preferences
(adapted from Zanardi et al.~\protect\cite{zanardi:2022}, Mendes Filho et al.~\protect\cite{filho:2017}, and Lee et al.~\protect\cite{lee:2025}). On the road, situations arise where a vehicle has to coordinate with other vehicles and must relax a low-priority metric to preserve more important metrics, such as giving way to an ambulance.}
\label{fig:scenario}
%\vspace{-1em}
\end{figure}

A preference is a ``total subjective comparative evaluation'' \cite{hausman:2011}.
In the context
of an autonomous vehicle, a total subjective comparative evaluation is a rule
that, given two trajectories, decides which
of the two is preferred from the perspective
of the controller.
For multiagent systems too,
preferences model the multiobjective
and often conflicting requirements
that the system must adhere to
for producing good behavior
as subjective comparative evaluations (\cref{fig:scenario}).

One way to express the tension between prioritized metrics
of multiple agents is to express them
as lexicographic optimization problems
in a game of ordered preference \cite{lee:2025}.
Games of ordered preference are multi-player games
with prohibitive computational cost.
Solving these games with a receding horizon alleviates this cost
through shorter time horizons and the introduction of feedback.
However, the receding horizon setting does not address the rapid dimensional increase caused
by adding players or preference levels.
Iterated best response~(IBR) efficiently solves multi-player games
by decomposing them into single-player problems,
where each player optimizes assuming fixed opponent strategies.
IBR convergence can be slow due
to the need to explore large opponent decision spaces.
We propose ``lexicographic IBR over time,''
an extension of IBR that incorporates past information,
available in receding horizon,
to make predictions about future opponent decisions
that accelerate convergence
to generalized Nash equilibria
for games of ordered preference.

To study this area, this paper asks:
\begin{enumerate}[itemsep=0pt,topsep=0pt,label=\textlangle ?\textrangle]
    \item \emph{%
        Can we efficiently compute approximate-optimal trajectories
        for games of ordered preference?}\label{Q1}
\end{enumerate}

To make progress on this question, we
\begin{enumerate}
\item find that lexicographic IBR over time approximates optimal solutions
for games of ordered preference; and
\item design experiments demonstrating that games
of ordered preference are an effective tool
for analyzing interactive preferences in transportation systems.
\end{enumerate}
\subsection{Related work}
\label{sec:org21f2b4e}

We have been primarily motivated
to study games of ordered preference
by Zanardi et al. \cite{zanardi:2022} and Lee et al. \cite{lee:2025}.
We explicitly build on this research
to compute for a receding horizon,
rather than a finite-time horizon,
to improve the scalability of these games.
We have additionally found insights in receding horizon games
\cite{hall:2022,hall:2024}, although the formulation is not directly applicable
to the problem of preference ordering.

Prior work, closely related to the formulation
of games of ordered preference in this paper,
is on \emph{prioritized metrics} 
and \emph{game-theoretic planning}.

\smallskip \noindent
\textsc{prioritized metrics} \quad 
Posetal games \cite{zanardi:2022} formalize games
with prioritized metrics
through partial orders over system designs \cite{censi:2019}, using symbolic structures to encode constraint and preference hierarchies \cite{wongpiromsarn:2021}.
While posetal games focus
on discrete decision spaces, we adopt lexicographic optimization \cite{penlington:2024},
that extends prioritized metrics to continuous action spaces \cite{lee:2025}.
However, the lexicographic nested problem structure introduces computational challenges: lower-priority objectives are optimized only after higher-priority ones are satisfied, creating a hierarchy of interdependent subproblems. To mitigate this computational intractability, prior work transcribes the lexicographic hierarchy into mathematical programs with complementarity constraints \cite{nurkanovic:2024}, with the addition of relaxation schemes \cite{lee:2025}.
Yet, coupled constraints cause scalability to remain limited.
By framing the problem
within a receding horizon setting, we exploit the temporal decoupling inherent
to IBR algorithms, which are known to be computational tractable.

\smallskip \noindent
\textsc{game-theoretic planning}\quad Game-theoretic traffic
management is needed to analyze the interactive nature of preferences between
autonomous agents \cite{qin:2024}. Efficient solution strategies exist
for game-theoretic planning \cite{laine:2021,williams:2023,mehr:2023,zhu:2023}, but solving games that involve nested optimization problems typically requires approximate solutions, similar to those used in solving trajectory games with continuous action spaces \cite{jond:2022,peters:2022,liu:2023}.
Moreover, predictions of other agents' intentions combined with planning
increase the accuracy of approximate solutions \cite{li:2022,zhang:2023}. Receding
horizon in game-theoretic planning has been studied in the classical formulation
\cite{veer:2023} and in variants---where objects are divided into immediate
and mid-term metrics \cite{fisac:2019}. These approaches do not consider the
problem of preferences in agent interactions.
\section{Preliminaries}
\label{sec:orgb84c929}

In this section, we review key concepts
in trajectory games, lexicographic minimization,
and games of ordered preference.

\smallskip \noindent
\textsc{notation} \quad For any non-negative number \(n\), we define the set
of integers \([n] \coloneq \{1, 2, \dots, n\}\).
We use boldface to represent a time-indexed vector with length
of a finite-time horizon game \(\mathbf{z} \coloneq \{z_0, z_1, \dots, z_{T_g-1}\}\).
We time index this vector between \(t_0 \in \mathbb{N}\) and \(t_1 \in \mathbb{N}\)
as the vector \(\mathbf{z}_{t_0:t_1} \coloneq \{z_{t_0}, z_{t_0+1}, \dots, z_{t_1}\}\).
We use superscripts as in \(\square^i\) to to denote the agent \(i\in[N]\).
Negation is used to include all agents but one,
\(\square^{-i} \coloneq \square \setminus \square^i\).
\subsection{Trajectory games}
\label{sec:org91af826}
\label{sec:trajgames}

Trajectory games are non-cooperative, multiagent,
general-sum constrained dynamic games.
Solutions to trajectory games amount to finding
generalized Nash equilibrium (GNE) in which all agents
adopt optimal discrete-time trajectories from which
none can unilaterally deviate without incurring an increase in cost.
In a trajectory game the state and action spaces for each agent \(i \in [N]\) are
defined as \(\mathcal{X}^i \subseteq \mathbb{R}^n\)
and \(\mathcal{U}^i \subseteq \mathbb{R}^m\) respectively,
with states usually encoding the position
of agents in the environment
and actions encoding their control inputs.
We assume that states and actions can be split by agent.
At discrete-time step \(t\), the game decision
of each agent, \(z_t^i = [x_t^i, u_t^i]\)
is composed by the \(i\text{th}\) agents' current state, \(x_t^i \in \mathcal{X}^i\),
and control input to be applied, \(u_t^i \in \mathcal{U}^i\), with the resulting
decision space being \(\mathcal{Z}^i \subseteq \mathcal{X}^i \times \mathcal{U}^i\).
We define the trajectory of each agent
as a sequence of decisions \(\mathbf{z}^i = \{z_0^i, \dots, z_{T_g-1}^i\}\)
over the time horizon of the game \(T_g\).

An individual cost function \(J^i \colon \mathcal{Z}^{i} \times \mathcal{Z}^{-i}  \to \mathbb{R}\)
encodes agent objectives over a trajectory,
as well as a set of private equality, \(g^i\),
and inequality, \(h^i\), constraints, with \(g^i \colon \mathcal{Z}^{i} \to \mathbb{R}\)
and \(h^i \colon \mathcal{Z}^{i} \to \mathbb{R}\).
Agents must also satisfy a set of shared equality, \(g^s\),
and inequality, \(h^s\) constraints,
that consider the global system state and for which all agents are equally responsible,
with \(g^s \colon \mathcal{Z}^{i} \times \mathcal{Z}^{-i} \to \mathbb{R}\)
and \(h^s \colon \mathcal{Z}^{i} \times \mathcal{Z}^{-i} \to \mathbb{R}\).
Solutions consist
of joint trajectories
for all the agents, \(\mathbf{z}\).

We can package the above as a generalized Nash equilibrium problem (GNEP) \cite{liu:2023}.
For each agent \(i\), the open loop, complete information GNEP is defined as
\begin{align}
\begin{split}
\minimize_{\mathbf{z}^i} \quad & J^i\left(\mathbf{z}^i, \mathbf{z}^{-i}; \theta^i \right) \\
\text{subject to} \quad & \eqnmarkbox[olive]{dynamics}{x_{t+1}^i=f^i\left(x_t^i, u_t^i\right)\; \text{for all} \; t \in[T_g-1]} \\
& \\ % Space for the annotation
& \eqnmarkbox[gray]{initial}{x_1^i=\hat{x}_1^i} \\
& g^i(\mathbf{z}) = 0 \\
& h^i(\mathbf{z}) \geq 0 \\
& \eqnmarkbox[brown]{shared1}{g^s\left(\mathbf{z}, \mathbf{z}^{-i}\right) = 0\!\:\:} \\[-.5em]
& \eqnmarkbox[brown]{shared2}{h^s\left(\mathbf{z}, \mathbf{z}^{-i}\right) \geq 0.}
\end{split}
\label{eq:trajgame}
\end{align}
\annotate[yshift=.5em]{above,right}{dynamics}{dynamics}
\annotate[xshift=1em,yshift=.5em]{right}{initial}{initial state}
\annotate[yshift=-.5em]{below}{shared2}{shared constraints\\e.g., collision avoidance}

\vspace{1.7em}\noindent
The objective function, additionally parametrized by \(\theta^i\), and constraints
take as arguments the trajectory \(\mathbf{z}\) of all agents,
differentiating the \(i\text{th}\) agent trajectory \(\mathbf{z}^i\) from
the others \(\mathbf{z}^{-i}\).
\subsection{Lexicographic minimization}
\label{sec:orgb3a4a9f}
\label{sec:lexmin}

For each \(k \in [K]\) let \(J_k \colon \mathbb{R}^n \to \mathbb{R}\)
be the \(k\text{th}\) objective function
of the decision variables residing
within a feasible set \(\mathbf{z} \in \mathcal{Z}\).
We define a total order \(\succcurlyeq_J\) on decision variables as follows:
for any \(\mathbf{z}, \mathbf{z}' \in \mathcal{Z}\),
we say that \(\mathbf{z} \succcurlyeq_J \mathbf{z}'\)
if and only if \(\mathbf{z} = \mathbf{z}'\)
or \(J_k(\mathbf{z}) < J_k(\mathbf{z}')\)
for the smallest \(k\) such that \(J_k(\mathbf{z}) \neq J_k(\mathbf{z}')\).
In other words, \(J_1\) holds the highest priority and \(J_K\) the lowest.

A lexicographic minimum of a feasible set \(\mathcal{Z} \in \mathbb{R}^K\)
is a decision variable \(\mathbf{z}^* \in \mathcal{Z}\)
for which \(\mathbf{z} \preccurlyeq_J \mathbf{z}^*\) for all \(\mathbf{z} \in \mathcal{Z}\).
Computing all lexicographic minima amounts
to a nested optimization problem.
The total order \(\succcurlyeq_J\) encodes a strict hierarchy, meaning
that given any two decision variables \(\mathbf{z}_h, \mathbf{z}_l\),
with \(h < l\), there exists a clear preference
for one over the other, \(\mathbf{z}_h \succcurlyeq_J \mathbf{z}_l\)
(we say \(\mathbf{z}_h\) has higher priority than \(\mathbf{z}_l\)),
and no two priorities are
at the same level of importance.
\subsection{Games of ordered preference}
\label{sec:org42f1672}
\label{sec:goop}

In a game of ordered preference agents pursue multiple,
hierarchically ranked objectives.
Unlike other games, where agents trade off objectives,
here agents resolve conflicts through lexicographic minimization,
strictly prioritizing goals. 

At each preference level \(k \in [K^i]\), agent \(i\) selects a trajectory \(\mathbf{z}_k^i\)
from a feasible set that is constrained by the lower (more prioritized) preference levels.
Agents know the final trajectories
of the other agents, \(\mathbf{z}^{-i}\),
when optimizing at level \(k\).
The game of ordered preference is equivalent
to a lexicographic minimization problem that computes a GNE,
where no agent can improve a higher-priority objective without
violating constraints or compromising lower-priority goals.
The solution to a game of ordered preference is found by jointly solving the following lexicographic minimization problem for each agent \(i\) across the \(K^i\) priority levels,
\begin{align}
\begin{split}
\minimize_{\mathbf{z}^i_{K^i}}   &\; J^i_{K^i}\left(\mathbf{z}^i_{K^i}, \mathbf{z}^{-i}; \theta^i \right) \\
\text{subject to} &\; \mathbf{z}^i_{K^i} \in \argmin_{\mathbf{z}^i_{K^i-1}}{J_{K^i-1}^i\left(\mathbf{z}^i_{K^i-1}, \mathbf{z}^{-i}; \theta^i\right)} \\[1em]
                  & \qquad \qquad \text{\rotatebox[origin=c]{150}{$\ddots$}} \\
                  &\; \text{subject to }\; \mathbf{z}^i_2 \in \argmin_{\mathbf{z}^i_1} \eqnmarkbox[olive]{p1}{J^i_1\left(\mathbf{z}^i_1, \mathbf{z}^{-i}; \theta^i \right)}\\
                  &\qquad \text{subject to } \mathbf{z}_1^i \in \mathbb{R}^{n}\\[1em]
                  &\qquad \qquad \text{such that } \eqnmarkbox[gray]{pc1}{g^i{\left(\mathbf{z}^i, \mathbf{z}^{-i}\right)} = 0,} \text{\;and}\\[-.5em]
                  & \phantom{\qquad \qquad \text{such that }}  \eqnmarkbox[gray]{pc2}{h^i{\left(\mathbf{z}^i, \mathbf{z}^{-i}\right)} \geq 0\hspace{.215em}}\\
                  &\; \text{such that } \eqnmarkbox[brown]{sc1}{g^s\left(\mathbf{z}\right) = 0,\hspace{0.065em}} \text{\;and}\\[-.5em]
                  &\; \phantom{\text{such that }} \eqnmarkbox[brown]{sc2}{h^s\left(\mathbf{z}\right) \geq 0.}
\label{eq:trajgoop}
\end{split}
\end{align}
\annotate[yshift=.5em]{above}{pc1}{private constraints\\i.e., feasibility region}
\annotate[yshift=-.5em]{below}{sc2}{shared constraints\\e.g., collision avoidance}
\annotate[yshift=.5em]{above,left}{p1}{principal preference\\e.g., speed limit}

\vspace{1.7em}\noindent

The optimal solution for each agent,
that is, the actual strategy that they will want to follow,
is the solution computed for the first level 
of the lexicographic minimization problem, \(\mathbf{z}_{K^i}^i\) in \cref{eq:trajgoop},
which embeds all the feasible set defining constraints
of inner levels.
For notational simplicity, we refer
to the solution
of the game
for each agent \(i\)
as \(\mathbf{z}^i \coloneq \mathbf{z}_{K^i}^i\).

\begin{figure*}[!t]
    \centering
\vspace{.5em}
    \begin{subfigure}[t]{0.38\linewidth}
        \def\svgwidth{\linewidth}
        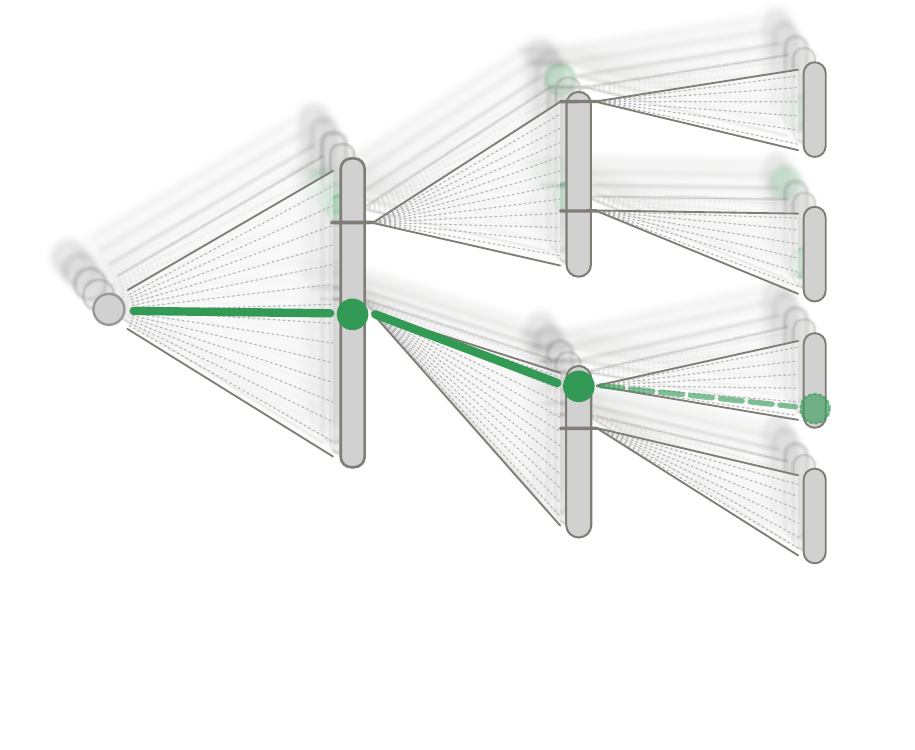
        \caption{First decision-making step.}
        \label{fig:game_graph_1}
    \end{subfigure}
    \hspace{\fill}
    \begin{subfigure}[t]{0.60\linewidth}
        \def\svgwidth{\linewidth}
        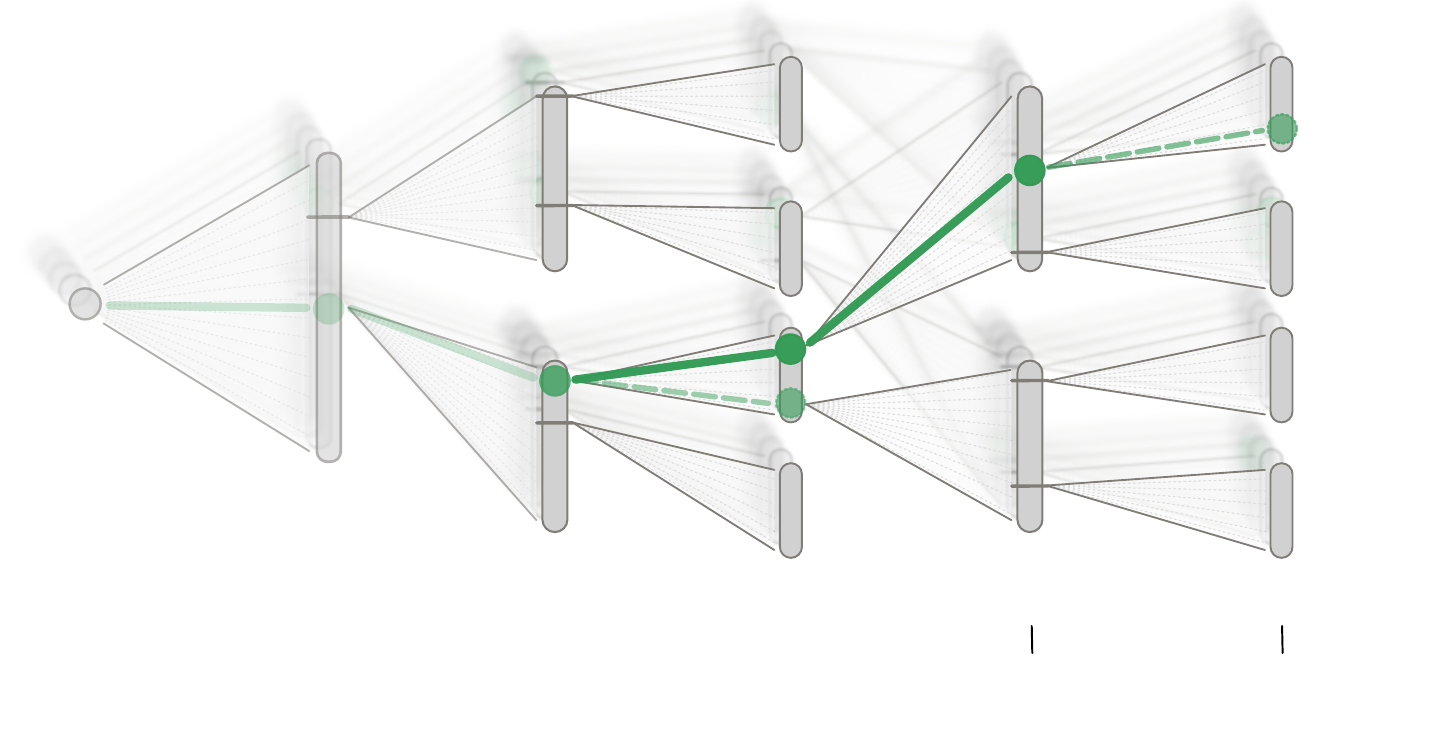
        \caption{Second decision-making step.}
        \label{fig:game_graph_2}
    \end{subfigure}
    
    \caption{
        In a trajectory game, each agent's strategy defines an optimal
        trajectory governed by a continuous state that evolves through actions
        selected from a continuous control input space at discrete time
        intervals. Agents compute a receding horizon to dynamically adapt their
        trajectories in response to environmental changes, periodically updating
        their strategies every $T_l$ time steps. At each decision stage, agents
        solve a game-theoretic optimization problem over a finite time horizon
        $T$, generating optimal trajectories (dashed lines). From these
        trajectories, agents execute only the first $T_l$ control inputs
        (solid lines) before resolving the game with updated environmental
        information.}
    \label{fig:game_graph}
\end{figure*}

In a trajectory game, the set of private constraints of an agent that restrict
their original feasibility regions typically include their dynamics and initial state.
The private constraints need to take into account the
preference structure of the other agents, making
\(g^i \colon \mathcal{Z}^{i} \to \mathbb{R}\)
and \(h^i \colon \mathcal{Z}^{i} \to \mathbb{R}\).
Additionally, a common shared constraint for these type of games
is collision avoidance (\cref{sec:trajgames}).
This enables
a more expressive definition of an agent's objectives, including those
that have utmost importance (such as safety constraints or following the speed limit)
and should only ever be suboptimal in the presence of a hard constraint.
For example, an autonomous vehicle should always prioritize not breaking
the speed limit (higher priority objective) even at the cost
of reaching their goal (lower priority objective) at a later time,
only ever speeding past it to avoid a collision (hard constraint).

It is often desirable to flatten the hierarchy of nested subproblems in \cref{eq:trajgoop} into a single level. For each agent, beginning with their innermost problem, we derive the Karush-Kuhn-Tucker (KKT) conditions and incorporate the resulting dual variables as induced primals in the outer problems. This effectively constrains the feasible set, ensuring that solutions conform to the lexicographic order of preferences. The resulting formulation is a mathematical program with complementarity constraints for each agent, which can be regularized by a relaxation scheme that expands the feasible set to solve the mixed complementarity problem (MCP) \cite{lee:2025}.
\section{Efficiently finding solutions to~games~of~ordered~preference}
\label{sec:orgb509d8a}

In this section, we formulate games of ordered preference into a receding horizon setting
and produce algorithms that exploit lexicographic IBR over time for efficient computation.
\subsection{Receding horizon games of ordered preference}
\label{sec:orgf03dc10}
\label{sec:rhgoop}

Games of ordered preference introduce considerable complexity due
to the nested structure of the optimization problem they generate.
Even when the hierarchy is reduced
to single-level problems, the dimensionality
of the decision variables increases substantially.
Solving trajectory games framed
in terms of dynamical systems within the context
of games of ordered preference is even more challenging due
to the higher dimensionality needed to represent the dynamics, along with spatiotemporal constraints
and time-indexed objectives, particularly when considering long, fixed-time horizons (\cref{sec:trajgames}).
The challenge of high dimensionality restricts their use
to instances with short, fixed-time horizons.
This limitation in turn diminishes agents' ability
to adapt to medium- and long-term environmental changes.

A receding horizon game of ordered preference mitigates this problem
by partitioning the solution space into manageable segments (\cref{fig:game_graph}).
In a receding horizon game
with length of a finite-time horizon, \(T_g\), agents periodically re-optimize their strategies.
At each decision-making stage, however, they focus on a reduced time horizon, \(T \ll T_g\),
repeatedly solving optimization subproblems
with the latest state observations serving
as the initial conditions \cite{christophersen:2007}.
At each decision-making stage, the agents extract and execute only the first \(T_l\) decisions
of the newly computed optimal joint strategy,
where \(0 < T_l \leq T\) is called
the \emph{turn length} \cite{peters:2025}.
These decision-making stages occur every \(T_l\) time steps,
continuing until the end of the game at time \(T_g\).

A receding horizon variation of a game
of ordered preference involves each rational agent \(i \in [N]\)
finding an optimal strategy
that lexicographically minimizes their ordered cost functions \(J^i\)
over the game time horizon \(t \in \{0, \dots, T_g-1\}\).
The optimal strategy along the entire game horizon \(T_g\),
which we denote as \(\mathbf{z}^{i,*}\), is incrementally built
by aggregating the executed steps of the partial solutions
obtained at each decision-making stage.

At each decision-making stage,
occurring every \(T_l\) time steps, a game
of ordered preferences is solved
over a reduced time horizon \(T\).
Each agent minimizes the partial cost function
$$\ell^i\left(\mathbf{z}^i_{t:t+T-1}, \mathbf{z}^{-i}_{t:t+T-1}; \theta^i\right)\text{.}$$
The partial cost \(\ell^i\) for each agent \(i\), with \(\ell^i \colon \mathcal{Z}^{i} \to \mathbb{R}\),
is the cost of executing the next \(T\) decisions \(\mathbf{z}^i_{t:t+T-1}\)
given the other agent's decisions \(\mathbf{z}^{-i}_{t:t+T-1}\).
It differs from the objective function \(J^i\) in \cref{eq:trajgoop} in that
the latter accounts for the entire game horizon \(T_g\), while \(\ell^i\) gives the cost for a
single reduced horizon solution for agent \(i\)
computed at time step \(t\),
\begin{equation}\label{eq:ji}
J^i\left(\mathbf{z}^i, \mathbf{z}^{-i}; \theta^i\right) = \sum_{t = 0}^{T_g-1} \ell^i\left(\mathbf{z}^i_{t:t+T-1}, \mathbf{z}^{-i}_{t:t+T-1}; \theta^i\right).
\end{equation}
We repeat \cref{eq:ji} at times \(t = m \cdot T_l\) for all positive integers \(m\) such that \(m \cdot T_l < T_g\).

The solution of a single decision-making stage at time step \(t\)
is a joint optimal strategy \(\mathbf{z}^*_{t:t+T-1}\).
Agents computing a receding horizon will find
at once their optimal strategy for the next \(T\) time steps,
\begin{equation*}
\mathbf{z}^{i,*}_{t:t+T-1} = \argmin_{\mathbf{z}^{i}_{t:t+T-1}} \ell^i \left(\mathbf{z}^i_{t:t+T-1}, \mathbf{z}^{-i,*}_{t:t+T-1}; \theta^i \right).
\end{equation*}

The optimal solution for each agent \(i\) at time step \(t\) corresponds
to the solution of the lowest priority level in~\cref{eq:trajgoop}.
After each decision-making stage, only the first \(T_l\) decisions
of the solution for all agents, \(\mathbf{z}^{*}_{t:t+T_l-1}\),
are aggregated into the final solution \(\mathbf{z}^*_{0:T_g-1}\),
with the associated control inputs
executed by the agents to evolve their states.
At step \(t+T_l\), a new decision-making stage takes place,
which solves the game again for \(t+T_l+T\),
using initial state \(x^i_{t+T_l}\).
This computation is repeated until the end of the game
is reached at time \(T_g\).
The algorithmic interpretation of this system involves
a measurement step at time \(t\) where the global state \(\mathbf{x}_t\) is observed
after evolving the system for \(T_l\) time steps following the previous decision-making
stage (\cref{alg:rhgoop}).

When comparing receding horizon formulations of games
against the corresponding fixed-time horizon ones, the benefits are two fold.
First, by solving for only \(T \ll T_g\) time steps at a time,
a receding horizon approach circumvents the computational
complexity associated with accounting for the entire solution at once.
Second, deferring the resolution of the final steps allows for adaptive
decision making in response to evolving environmental conditions,
allowing agents to reconsider their strategies over time (\cref{fig:game_graph}).
\subsection{Lexicographic IBR over time}
\label{sec:org2c771d8}

Games of ordered preference
incur a significant increase
in the problem's dimensions for each preference level.
The significant increase arises from the accumulation
of extra equality and inequality constraints and induced primals
derived from the KKT conditions at each level
during the flattening of the nested optimization problem \cite{lee:2025}.
The flattening of multiple preference levels, while necessary to solve games
of ordered preference, can quickly grow the problem dimensions
to become computationally intractable.
One way to mitigate the problem of high dimensionality
is to partition the coupled multi-player game of ordered preference
into smaller single-player subproblems
which can be solved with extensively used algorithms such as IBR.

\begin{algorithm}[!t]
\KwData{MCP for the game of ordered preference, game~horizon $T_g$, receding~horizon $T$, turn~length $T_l$, and initial states $\mathbf{x}_0$}
\KwResult{Optimal joint strategies $\mathbf{z}^{*}$}
\medskip
Start at $t = 0$ with initial states $\mathbf{x}_0$ for each agent~$i$\;
\While{$t < T_g$}{
    Measure global state $\mathbf{x}_{t}$ at time $t$\;
    Solve MCP at $\mathbf{x}_t$ with time horizon $T$\;
    Evolve state for time $T_l$ using $\mathbf{z}_{t:t+T_l-1}^*$\;
    Recede horizon: $t \gets t + T_l$\;
}
$\mathbf{z}^{*} \gets \mathbf{z}_{0:T_g-1}^{i,*}$ for each agent $i$\;
\caption{Receding horizon for games of ordered preference.}
\label{alg:rhgoop}
\end{algorithm}

In the standard formulation of IBR, multiple agents repeatedly take turns to
compute their best response considering other players' strategies.
Under certain assumption that are true for games of ordered preference
IBR converges to a Nash equilibrium \cite{fridovich-keil:2024}.
IBR in the receding horizon setting
uses information about past decisions of other agents
to better predict their future best responses, termed ``IBR over time.''

In lexicographic IBR over time, the multiagent problem is split in \(N\)
single-player games, with each player performing lexicographic
optimization of their preferences while considering other players'
strategies as fixed.
The aggregate solution of the single-player games
is an approximation of the GNE
described in the multiagent formulation.
Lexicographic IBR over time differs from IBR by starting with a more accurate initial guess of the
trajectories of other agents and stopping at a fixed number of iterations.
In particular, IBR over time improves standard IBR efficiency
by warmstarting agents' best responses, which reduces the number of iterations needed
to converge to an equilibrium.

Approximate solutions are acceptable when they are close to the optimum.
Lexicographic IBR over time produces approximate-optimal solutions, provided there are good predictions
of other agents' strategies and an acceptable number of iterations.
An effective strategy for generating good predictions is
to use the trajectories of other agents computed
in the previous decision-making stage.
Further refinement occurs in additional IBR iterations, where each agent sequentially considers the running best responses
of others until convergence to the GNE is achieved or a maximum number of iterations (\(L\) in \cref{alg:ibr}) is reached.

The complete predictions for the first iteration of IBR at a decision-making stage
are obtained by evolving the current measured state
using the dynamics with the action sequences,

\vspace{1em}
$$
z^i_{t+1} = \left[x^i_{t+1}, u^i_{t+1}\right] \approx \left[\eqnmarkbox[olive]{dyn}{f^i\big(x^i_t, u^i_t\big)}, \eqnmarkbox[gray]{ut}{u^i_t}\right].
$$
\annotate[yshift=.5em]{above,left}{ut}{preceding action}
\annotate[yshift=-.5em]{below}{dyn}{dynamics}
\vspace{.75em}

\begin{figure}[!t]
\def\svgwidth{\linewidth}
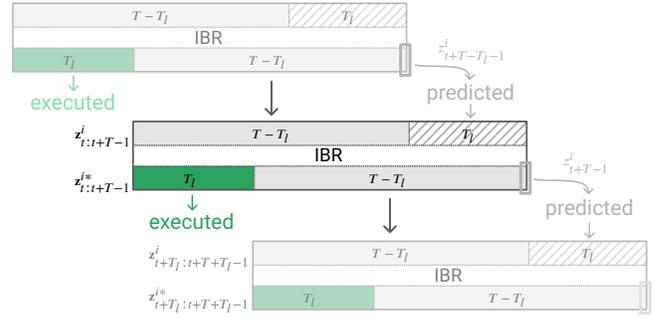
\caption{
    When solving using IBR, the predictions of the other agents' trajectories
    are based on the solutions obtained at the last decision-making stage.
    The previous solutions are shifted in time and then padded with additional action,
    which can be a null action $u_\varnothing^i$ or the same action as the previous step $u_{t-1}^i$.
    }
\label{fig:time-padding}
\end{figure}

\noindent
The action sequences are obtained by shifting those of previously computed trajectories
and padding them with additional actions to fill the time horizon \(T\) (\cref{fig:time-padding}).

In IBR, the lexicographic minimization of each player's objectives
can be achieved by successively solving single-player optimization problems.
For each each preference level \(k\) of player \(i\), we solve a minimization problem
with cost function \(\ell^i_k\) as its objective. In this setting, the other
player's trajectories become another parameter of the game, included in \(\theta^i\).
The optimal cost value computed at any level
\(y_k^{i,*} = \ell^i_k(\mathbf{z}^i_{k}; \theta^i)\) is recorded.
Then, in each subsequent outer level, we add additional inequality constraints
to preserve the optimality with respect to
previously optimized inner objectives \cite{kochenderfer:2019}.
For any level \(k\), the added constraints are
$$\ell^i_j\left(\mathbf{z}^i_{k}; \theta^i\right) \leq y_j^{i,*}$$
for each previous level \(1 \leq j < k\).
Given that the optimization method used is not optimal
since it is based on iterated approximations,
by using inequality constraints we ensure that if in subsequent optimizations
better solutions for previous levels' objectives are encountered they won't be rejected.

The IBR over time solution, described in \cref{alg:ibr}, acts as a replacement
of the MCP solving step in \cref{alg:rhgoop}
in the single-player formulation.

In contrast to the single, coupled optimization problem solved
for the game of ordered preference, IBR over time solves
a total of \(\sum_{i=1}^{N} \abs{K^i}\) problems at each decision-making stage,
where \(K^i\) represents the number of preference levels for agent \(i\).
However, the reduced dimensionality of the individual problems results
in significantly faster computation times.
Moreover, the IBR over time solution can be computed in parallel for each agent \(i\),
rather than sequentially.
This parallelization can further enhance processing efficiency, particularly in games with many agents.
\section{Experiments}
\label{sec:org3c0c378}

In this section we evaluate the performance of IBR over time
for solving games of ordered preference. 
We design experiments to (1) examine how different preference relations
yield qualitatively different solutions for the same scenario
and (2) quantitatively compare the efficiency of IBR over time
against the standard formulation of games of ordered preference
in receding horizon.

\begin{algorithm}[!t]
\KwData{MCP for the game of ordered preference, current~time $t$,
    preceding trajectories $\mathbf{z}^{i}_{t:t+T-T_l-1}$,
    time~horizon $T$, maximum iterations $L$, and~convergence tolerance $\epsilon$}
\KwResult{Approximate-optimal joint strategies $\mathbf{z}^{*}_{t:t+T-1}$}
\medskip
$\mathbf{z}_{0:T-1} \gets$ Shift and pad preceding trajectories\;
Best responses $\gets \mathbf{z}_{0:T-1}$\;
\For{$1$ \KwTo $L$}{
    \For{$i = 1$ \KwTo $N$}{
        Parameters $\theta^i$ $\gets$ best response of agents $-i$\;
        $\mathbf{z}^{i}_{0:T-1} \gets \argmin_{\mathbf{z}^i_{0:T-1}} \ell^i(\mathbf{z}^{i}_{0:T-1}; \theta^i)$\;
        Update best response of agent $i$\;
    }
    \If{$\text{solution improvement} < \epsilon$}{
        \textsc{break}\;
    }
}
% \Return{$\mathbf{z}^{*}_{t:t+T-1}\coloneq\mathbf{z}_{0:T-1}$}
\caption{A single IBR solution in games of ordered preference.}
\label{alg:ibr}
\end{algorithm}
\subsection{Implementation details}
\label{sec:org507a744}

We implement the code
in Julia \cite{julia}, build upon the code
of Lee et al. \cite{lee:2025},
and make extensive use of TrajectoryGamesBase.jl \cite{peters:2025}
to instantiate and solve games
of ordered preference
in the receding horizon setting.
The benchmarks use the package ParametricMCPs.jl \cite{peters:2024},
which is a wrapper around the PATH solver \cite{dirkse:1995}.
We package the code
for running the following experiments as LEXIBROverTime.jl \cite{code}.

All benchmarks are run on a single core
of a machine with no additional load.
\subsection{Qualitative evaluation of approximate solutions}
\label{sec:org5025c01}

\begin{figure*}[!t]
\centering
    \begin{subfigure}[t]{0.49\linewidth}
        \def\svgwidth{\linewidth}
        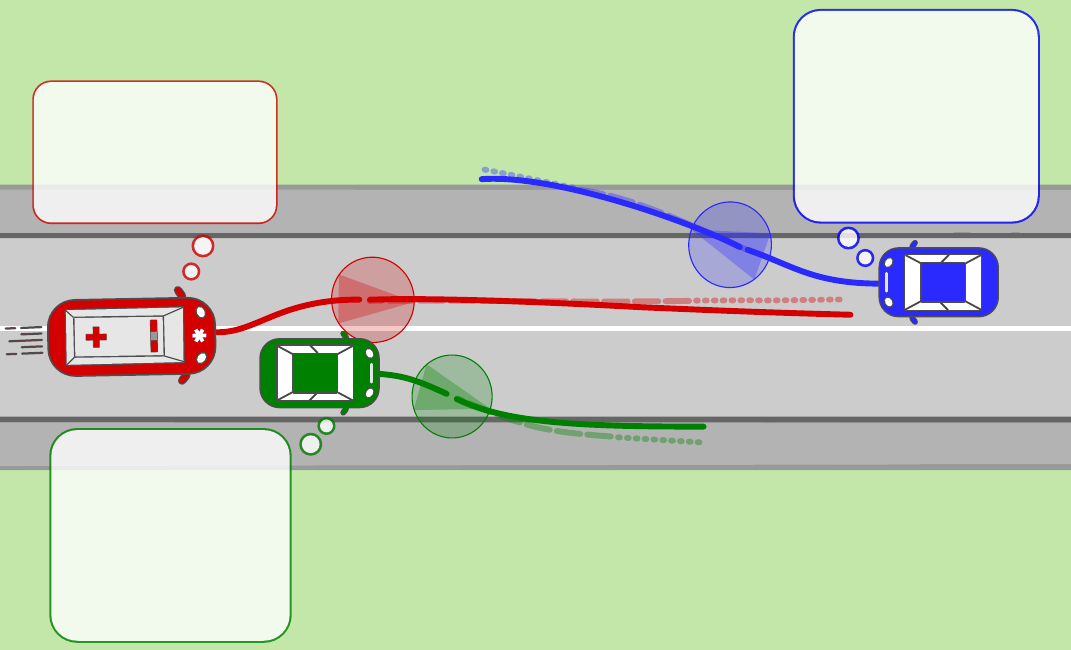
        \caption{Highway.}
        \label{fig:field_scenario}
    \end{subfigure}
    \hspace{\fill}
    \begin{subfigure}[t]{0.49\linewidth}
        \def\svgwidth{\linewidth}
        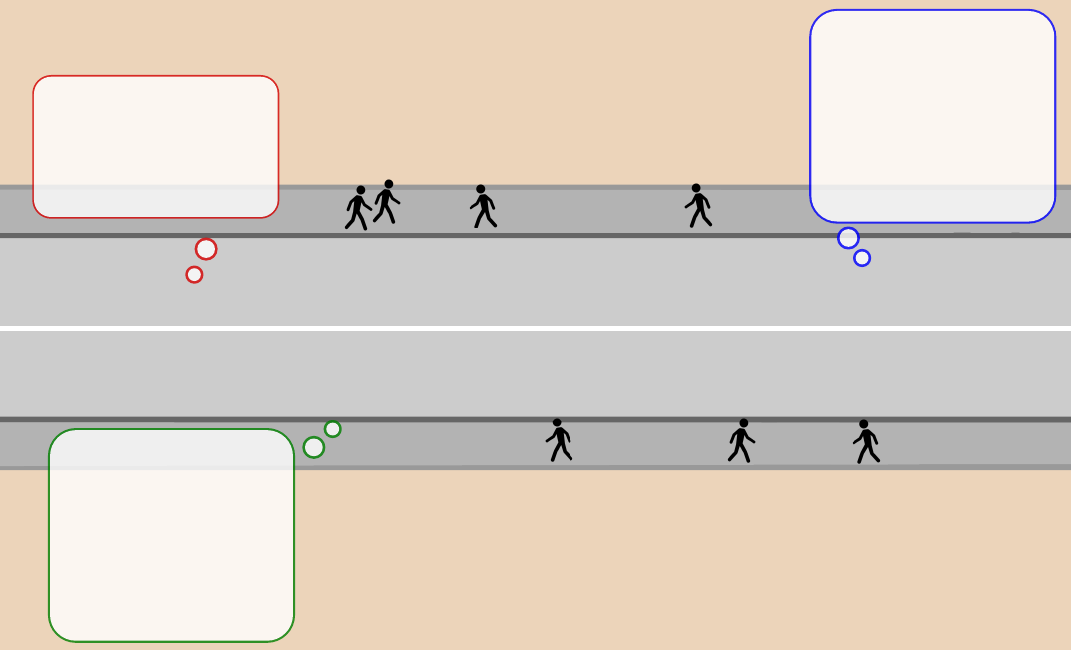
        \caption{City.}
        \label{fig:city_scenario}
    \end{subfigure}
\caption{
    Different preference relations model varying contexts, resulting in qualitatively different agent behaviors.
    In a highway scenario, vehicles prefer leaving their lane to avoid braking hard because there are no pedestrians,
    while in an urban scenario they prefer to stay within the road to not risk hitting pedestrians.
    Dashed lines show each agent's trajectory as predicted by other agents.
    Solid lines represent their actual computed trajectories.
    }
\label{fig:road_scenario}
\end{figure*}

\noindent
\textsc{evaluation scenario} \quad
To qualitatively assess
how different preference relations influence the solution
of a game of ordered preference, given identical parameters,
we consider a road navigation problem involving three players:
two cars driving in opposite lanes
of a road and an ambulance attempting to rush past them
in response to an emergency.
We present two simulation instances
with the same parameters (road layout, initial state, dynamics, constraints, etc.), but with different preference relations (\cref{fig:road_scenario}).
The game is solved using IBR over time with a single IBR iteration (\(L=1\))
and the same time horizon, turn length,
and simulation times for both scenarios.

\smallskip \noindent
\textsc{main result} \Circled{\oldstylenums 1} -- \textsc{ibr over time distinguishes different preference relations} \quad
Trajectories of agents
with different preference relations vary (\cref{fig:road_scenario}).
In the highway scenario, the green car slightly veers out
of its lane to allow the ambulance to pass,
while the blue car temporarily leaves the road
to avoid a potential collision.
In contrast, in the urban scenario, the green car remains
at the edge of its lane,
and the blue car executes a sharp turn towards the center
of the road to avoid pedestrians,
while the ambulance overtakes in the opposite lane.
This demonstrates how IBR over time identifies solutions
that respect the preference hierarchy
of the agents, thereby replicating the outcomes derived
from the standard formulation.

\begin{table}[!t]
\centering
\caption{Performance comparison between baseline
         and IBR over time with different iteration limits $L$
         and number of preference levels $K$.
         Times and $L_1$ distances are averaged across runs.
         }
\label{tab:results}
\begin{tabular}{@{}lrrrr@{}}
& \multicolumn{2}{c}{$K=2$} & \multicolumn{2}{c}{$K=3$} \\
\cmidrule(r){2-3}\cmidrule(l){4-5}
Method   & $t_\text{solve}$     & $L_1$ distance & $t_\text{solve}$     & $L_1$ distance \\ \midrule
Baseline & $44.51 \si{\second}$ & --               & $2676 \si{\second}$  & --               \\ % TODO
$L=1$    & $0.36 \si{\second}$  & $3.28\times 10^{-4}$         & $0.32 \si{\second}$  & $5.45\times 10^{-4}$         \\
$L=2$    & $0.59 \si{\second}$  & $1.22 \times 10^{-5}$         & $0.63 \si{\second}$  & $6.56\times 10^{-5}$         \\
$L=3$    & $0.85 \si{\second}$  & $4.64\times 10^{-6}$         & $0.75 \si{\second}$   & $5.66\times 10^{-5}$         \\
$L=5$    & $1.39 \si{\second}$  & $1.78\times 10^{-6}$         & $1.20 \si{\second}$  & $4.96\times 10^{-5}$         \\
$L=10$   & $2.67 \si{\second}$  & $1.71\times 10^{-6}$         & $2.32 \si{\second}$  & $3.03\times 10^{-5}$         \\
$L=20$   & $5.40 \si{\second}$  & $1.28\times 10^{-6}$        & $4.55 \si{\second}$  & $8.51\times 10^{-5}$         \\
$L=1000$ & $269.4 \si{\second}$ & $2.45\times 10^{-6}$          & $273.27 \si{\second}$ & $4.14\times 10^{-5}$         \\
\bottomrule
\end{tabular}
\end{table}

\subsection{Efficiency of lexicographic IBR over time}
\label{sec:org3ead184}

\noindent
\textsc{evaluation scenario} \quad
To assess the efficiency of IBR over time
we consider a simplified variant of the previous problem:
two cars driving along parallel lanes
of a road in the same direction,
with an ambulance attempting to overtake them.
The fact that all vehicles are traveling
in the same direction reduces the risk of collision, resulting
in less restrictive optimization problems
that can typically be solved
using both IBR over time and the baseline method.
We conduct a Monte Carlo study with \(20\) variations
of the scenario, generated by adding random permutations sampled
from a uniform distribution centered
around the initial states (position and speed) of the agents.
The experiments are run with two different numbers
of preference levels \(K\), common to all players,
in order to account for the increased complexity
of the optimization problems associated
with a higher number of priority levels.

\smallskip \noindent
\textsc{baseline} \quad
For the baseline, we use an adaptation
of the solver for games of ordered preference
in their standard formulation \cite{lee:2025},
which operates in the receding horizon setting.
We use the same hyperparameters in both methods,
including the relaxation iterations specified by Lee et al. \cite{lee:2025}.

\smallskip \noindent
\textsc{evaluation metrics} \quad
The primary evaluation metrics
for performance comparison are the average time required
to compile the problems \(t_\text{compile}\) and the average time taken
to solve each of them \(t_\text{solve}\).
The compile time is a one-time overhead not included in the solution time.
For comparison, the solution times for IBR
over time are expressed as a percentage
of the baseline solution time.
Since there might be a continuum of equilibria, that is, multiple optimal solutions, we will evaluate the quality of the solutions by
computing the \(L_1\) distance between the solution trajectory and what would be the solution at the next IBR iteration \(L+1\).

\smallskip \noindent
\textsc{main result} \Circled{\oldstylenums 2} -- \textsc{ibr over time efficiently solves games of ordered preference} \quad
The results demonstrate that IBR outperforms the baseline
in both problem compilation and solution times,
with a substantial margin, even
as the number of iterations \(L\) increases (\cref{tab:results}).
The differences in solution times become more pronounced
as the number of preference levels increases,
with the baseline showing a much larger degradation in performance.
The average solution time for IBR is directly proportional
to the number of IBR iterations. A maximum number
of iterations can be adjusted to balance the optimality
of solutions with computational constraints.
The problem compilation time is not included in the results,
summarized in \cref{tab:results},
given that its value does not depend on the number of iterations \(L\).
The average baseline compilation times \(t_\text{compile}\) are \(1.50\si{\second}\) and \(9.37\si{\second}\) for preference levels \(K = 2\) and \(K = 3\) respectively.
For the same preference levels, IBR over time produces average compilation times
that are significantly lower, \(0.75\si{\second}\) and
\(1.10\si{\second}\),
primarily due to the reduced size of the problems
and increased solver efficiency.
The compilation of all subproblems in IBR over time takes
less total time than the compilation of a single problem in the baseline.
Shorter compilation times enable faster transitions
between preference relations,
which are embedded in the problem's structure.
Additionally, the \(L_1\) distance of the solution at iteration \(L\) with respect to
the next iteration \(L+1\) decreases rapidly
within a few iterations, indicating that the trajectories stabilize and become approximate-optimal.
The final average \(L_1\) distance does not reach zero (\cref{tab:results}).
This phenomenon arises because IBR does not converge to a stable equilibrium
for some problems.
The limit, \(L\), ensures the algorithm terminates in those cases.
\subsection{Discussion}
\label{sec:org9855716}

The experimental results demonstrate
that it is possible to efficiently find approximate solutions
to games of ordered preference (answering question \ref{Q1}).
Lexicographic IBR over time enables tractable computation
of single-player versions of games
of ordered preference while still respecting the agents' preference hierarchy.
Although the solutions are not guaranteed
to be optimal compared to the standard formulation, the number of IBR iterations can be adjusted
to reduce the gap to the optimal solution, provided time and computational resources are not constrained.
However, practical results indicate that even with a single IBR iteration, using previously computed trajectories as a base for predicting the other agents' trajectories provides a reasonable approximation.

The benchmarks demonstrate
that IBR over time reduces solution times
for games of ordered preference compared to existing methods, especially
as the number of preference levels increases.
This improved computational tractability enables solving larger problems
with more preferences and longer time horizons---both crucial factors
for developing agents capable
of complex behaviors and effective adaptation
to short- and long-term environmental changes.
However, the lexicographic structure of games of ordered preference can lead
to dominant strategies, where agents primarily optimize
for higher-priority objectives.
This tendency toward dominant strategies can be mitigated through careful modeling
of agents' preferences, ensuring
that the feasible regions defined
by successive preference levels still allow
for exploring lower-priority objectives.
The results also show that the solution stabilizes, that is, the solution changes less with respect to the previous iteration \(L-1\), as
number of iterations \(L\) grows. This stability of solutions means that with just a few iterations we can achieve
approximated-optimal solutions.
\section{Conclusion}
\label{sec:orgb53bdbf}

Autonomous vehicles are objective decision makers,
yet the quality of those decisions must be measured against the subjective,
context-aware choices of human drivers.
Preference relations model prioritized metrics
in multi-objective optimization,
which capture the subjective decision making of human drivers.
Accounting for the context-varying preferences and intentions
of all vehicles in the road
is a necessary condition for intelligent traffic management.
Games of ordered preference have demonstrated compelling support
for addressing this requirement.
However, the computational cost of considering preference hierarchies
has limited their inclusion in real-time systems' design.
Beyond real-time trajectory optimization, autonomous navigation
also demands that agents are able to react to the changing environments.
Our proposed algorithm, lexicographic IBR over time,
produces approximate-optimal solutions
to games of ordered preference in receding horizon
fast enough to potentially be practical.
Lexicographic IBR over time equips autonomous agents
in complex traffic scenarios with the ability
to adapt to the environment
in real time while ranking their preferences and considering those of others.
\section{Acknowledgments}
\label{sec:orgc7dbcaa}

P. de las Heras Molins and E. Roy-Almonacid are partially supported by the E.U.
Digital Europe Programme under grant DIGITAL-101083531-MERIT.

D. H. Lee and D. Fridovich-Keil are partially supported by the U.S. National
Science Foundation (NSF) under Grant No. 2336840.

G. Bakirtzis is partially supported by the academic and research chair
\emph{Architecture des Systèmes Complexes} through the following partners: Dassault Aviation, Naval
Group, Dassault Systèmes, KNDS France, Agence de l'Innovation de Défense, and
Institut Polytechnique de Paris.

\IEEEtriggeratref{12}
\printbibliography
\end{document}

%% file: figures/scenario.pdf_tex
%% Creator: Inkscape 1.4.2 (ebf0e940d0, 2025-05-08), www.inkscape.org
%% PDF/EPS/PS + LaTeX output extension by Johan Engelen, 2010
%% Accompanies image file 'scenario.pdf' (pdf, eps, ps)
%%
%% To include the image in your LaTeX document, write
%%   \input{<filename>.pdf_tex}
%%  instead of
%%   \includegraphics{<filename>.pdf}
%% To scale the image, write
%%   \def\svgwidth{<desired width>}
%%   \input{<filename>.pdf_tex}
%%  instead of
%%   \includegraphics[width=<desired width>]{<filename>.pdf}
%%
%% Images with a different path to the parent latex file can
%% be accessed with the `import' package (which may need to be
%% installed) using
%%   \usepackage{import}
%% in the preamble, and then including the image with
%%   \import{<path to file>}{<filename>.pdf_tex}
%% Alternatively, one can specify
%%   \graphicspath{{<path to file>/}}
%% 
%% For more information, please see info/svg-inkscape on CTAN:
%%   http://tug.ctan.org/tex-archive/info/svg-inkscape
%%
\begingroup%
  \makeatletter%
  \providecommand\color[2][]{%
    \errmessage{(Inkscape) Color is used for the text in Inkscape, but the package 'color.sty' is not loaded}%
    \renewcommand\color[2][]{}%
  }%
  \providecommand\transparent[1]{%
    \errmessage{(Inkscape) Transparency is used (non-zero) for the text in Inkscape, but the package 'transparent.sty' is not loaded}%
    \renewcommand\transparent[1]{}%
  }%
  \providecommand\rotatebox[2]{#2}%
  \newcommand*\fsize{\dimexpr\f@size pt\relax}%
  \newcommand*\lineheight[1]{\fontsize{\fsize}{#1\fsize}\selectfont}%
  \ifx\svgwidth\undefined%
    \setlength{\unitlength}{453.54330709bp}%
    \ifx\svgscale\undefined%
      \relax%
    \else%
      \setlength{\unitlength}{\unitlength * \real{\svgscale}}%
    \fi%
  \else%
    \setlength{\unitlength}{\svgwidth}%
  \fi%
  \global\let\svgwidth\undefined%
  \global\let\svgscale\undefined%
  \makeatother%
  \begin{picture}(1,0.75)%
    \lineheight{1}%
    \setlength\tabcolsep{0pt}%
    \put(0,0){\includegraphics[width=\unitlength,page=1]{scenario.pdf}}%
    \put(0.45313058,0.65606461){\color[rgb]{0,0,0}\transparent{0.80000001}\makebox(0,0)[lt]{\lineheight{1.25}\smash{\begin{tabular}[t]{l}\small{$\wedge$}\end{tabular}}}}%
    \put(0.40340004,0.61561741){\color[rgb]{0,0,0}\transparent{0.80000001}\makebox(0,0)[lt]{\lineheight{1.25}\smash{\begin{tabular}[t]{l}\textsf{\small comfort}\\\end{tabular}}}}%
    \put(0.40299797,0.69234216){\color[rgb]{0,0,0}\transparent{0.80000001}\makebox(0,0)[lt]{\lineheight{1.25}\smash{\begin{tabular}[t]{l}\textsf{\small collision}\\\end{tabular}}}}%
    \put(0.44651599,0.57933986){\color[rgb]{0,0,0}\transparent{0.80000001}\makebox(0,0)[lt]{\lineheight{1.25}\smash{\begin{tabular}[t]{l}\small{$\succcurlyeq_J$}\end{tabular}}}}%
    \put(0.43342085,0.53623457){\color[rgb]{0,0,0}\transparent{0.80000001}\makebox(0,0)[lt]{\lineheight{1.25}\smash{\begin{tabular}[t]{l}\textsf{\small time}\\\end{tabular}}}}%
    \put(0,0){\includegraphics[width=\unitlength,page=2]{scenario.pdf}}%
    \put(0.6785759,0.65606459){\color[rgb]{0,0,0}\transparent{0.80000001}\makebox(0,0)[lt]{\lineheight{1.25}\smash{\begin{tabular}[t]{l}\small{$\wedge$}\end{tabular}}}}%
    \put(0.6288455,0.53624244){\color[rgb]{0,0,0}\transparent{0.80000001}\makebox(0,0)[lt]{\lineheight{1.25}\smash{\begin{tabular}[t]{l}\textsf{\small comfort}\\\end{tabular}}}}%
    \put(0.64076648,0.61369915){\color[rgb]{0,0,0}\transparent{0.80000001}\makebox(0,0)[lt]{\lineheight{1.25}\smash{\begin{tabular}[t]{l}\textsf{\small speed}\\\end{tabular}}}}%
    \put(0.62844329,0.69234216){\color[rgb]{0,0,0}\transparent{0.80000001}\makebox(0,0)[lt]{\lineheight{1.25}\smash{\begin{tabular}[t]{l}\textsf{\small collision}\\\end{tabular}}}}%
    \put(0.67196131,0.57933985){\color[rgb]{0,0,0}\transparent{0.80000001}\makebox(0,0)[lt]{\lineheight{1.25}\smash{\begin{tabular}[t]{l}\small{$\succcurlyeq_J$}\end{tabular}}}}%
    \put(0,0){\includegraphics[width=\unitlength,page=3]{scenario.pdf}}%
    \put(0.78189321,0.2504005){\color[rgb]{0,0,0}\transparent{0.80000001}\makebox(0,0)[lt]{\lineheight{1.25}\smash{\begin{tabular}[t]{l}\small{$\wedge$}\end{tabular}}}}%
    \put(0.73216267,0.13057833){\color[rgb]{0,0,0}\transparent{0.80000001}\makebox(0,0)[lt]{\lineheight{1.25}\smash{\begin{tabular}[t]{l}\textsf{\small comfort}\\\end{tabular}}}}%
    \put(0.76062002,0.20803504){\color[rgb]{0,0,0}\transparent{0.80000001}\makebox(0,0)[lt]{\lineheight{1.25}\smash{\begin{tabular}[t]{l}\textsf{\small time}\\\end{tabular}}}}%
    \put(0.73176055,0.28667803){\color[rgb]{0,0,0}\transparent{0.80000001}\makebox(0,0)[lt]{\lineheight{1.25}\smash{\begin{tabular}[t]{l}\textsf{\small collision}\\\end{tabular}}}}%
    \put(0.77527871,0.17367573){\color[rgb]{0,0,0}\transparent{0.80000001}\makebox(0,0)[lt]{\lineheight{1.25}\smash{\begin{tabular}[t]{l}\small{$\succcurlyeq_J$}\end{tabular}}}}%
  \end{picture}%
\endgroup%

%% file: figures/game_graph_1.pdf_tex
%% Creator: Inkscape 1.2.2 (b0a8486541, 2022-12-01), www.inkscape.org
%% PDF/EPS/PS + LaTeX output extension by Johan Engelen, 2010
%% Accompanies image file 'game_graph_1.pdf' (pdf, eps, ps)
%%
%% To include the image in your LaTeX document, write
%%   \input{<filename>.pdf_tex}
%%  instead of
%%   \includegraphics{<filename>.pdf}
%% To scale the image, write
%%   \def\svgwidth{<desired width>}
%%   \input{<filename>.pdf_tex}
%%  instead of
%%   \includegraphics[width=<desired width>]{<filename>.pdf}
%%
%% Images with a different path to the parent latex file can
%% be accessed with the `import' package (which may need to be
%% installed) using
%%   \usepackage{import}
%% in the preamble, and then including the image with
%%   \import{<path to file>}{<filename>.pdf_tex}
%% Alternatively, one can specify
%%   \graphicspath{{<path to file>/}}
%% 
%% For more information, please see info/svg-inkscape on CTAN:
%%   http://tug.ctan.org/tex-archive/info/svg-inkscape
%%
\begingroup%
  \makeatletter%
  \providecommand\color[2][]{%
    \errmessage{(Inkscape) Color is used for the text in Inkscape, but the package 'color.sty' is not loaded}%
    \renewcommand\color[2][]{}%
  }%
  \providecommand\transparent[1]{%
    \errmessage{(Inkscape) Transparency is used (non-zero) for the text in Inkscape, but the package 'transparent.sty' is not loaded}%
    \renewcommand\transparent[1]{}%
  }%
  \providecommand\rotatebox[2]{#2}%
  \newcommand*\fsize{\dimexpr\f@size pt\relax}%
  \newcommand*\lineheight[1]{\fontsize{\fsize}{#1\fsize}\selectfont}%
  \ifx\svgwidth\undefined%
    \setlength{\unitlength}{438.75bp}%
    \ifx\svgscale\undefined%
      \relax%
    \else%
      \setlength{\unitlength}{\unitlength * \real{\svgscale}}%
    \fi%
  \else%
    \setlength{\unitlength}{\svgwidth}%
  \fi%
  \global\let\svgwidth\undefined%
  \global\let\svgscale\undefined%
  \makeatother%
  \begin{picture}(1,0.82051282)%
    \lineheight{1}%
    \setlength\tabcolsep{0pt}%
    \put(0,0){\includegraphics[width=\unitlength,page=1]{game_graph_1.pdf}}%
    \put(0.82722177,0.2033719){\color[rgb]{0,0,0}\makebox(0,0)[rt]{\lineheight{1.25}\smash{\begin{tabular}[t]{r}$\mathbf{z}^{i, *}_{0:T-1}$\end{tabular}}}}%
    \put(0,0){\includegraphics[width=\unitlength,page=2]{game_graph_1.pdf}}%
    \put(0.11670402,0.05132989){\color[rgb]{0,0,0}\makebox(0,0)[t]{\lineheight{1.25}\smash{\begin{tabular}[t]{c}$0$\end{tabular}}}}%
    \put(0.63302159,0.05329505){\color[rgb]{0,0,0}\makebox(0,0)[t]{\lineheight{1.25}\smash{\begin{tabular}[t]{c}$T_l$\end{tabular}}}}%
    \put(0.89118038,0.05132989){\color[rgb]{0,0,0}\makebox(0,0)[t]{\lineheight{1.25}\smash{\begin{tabular}[t]{c}$T$\end{tabular}}}}%
    \put(0.37945997,0.05204362){\color[rgb]{0,0,0}\makebox(0,0)[t]{\lineheight{1.25}\smash{\begin{tabular}[t]{c}$\dots$\end{tabular}}}}%
    \put(0.48765165,0.24787638){\color[rgb]{0,0,0}\makebox(0,0)[t]{\lineheight{1.25}\smash{\begin{tabular}[t]{c}$x^{i}_{t}$\end{tabular}}}}%
    \put(0,0){\includegraphics[width=\unitlength,page=3]{game_graph_1.pdf}}%
    \put(0.28698669,0.30671875){\color[rgb]{0,0,0}\makebox(0,0)[rt]{\lineheight{1.25}\smash{\begin{tabular}[t]{r}$\mathbf{z}^{i, *}_{0:T_l}$\end{tabular}}}}%
    \put(0,0){\includegraphics[width=\unitlength,page=4]{game_graph_1.pdf}}%
    \put(0.22057496,0.71614975){\color[rgb]{0,0,0}\makebox(0,0)[rt]{\lineheight{1.25}\smash{\begin{tabular}[t]{r}\textsf{\small{player}} $i$\end{tabular}}}}%
    \put(0,0){\includegraphics[width=\unitlength,page=5]{game_graph_1.pdf}}%
  \end{picture}%
\endgroup%

%% file: figures/game_graph_2.pdf_tex
%% Creator: Inkscape 1.2.2 (b0a8486541, 2022-12-01), www.inkscape.org
%% PDF/EPS/PS + LaTeX output extension by Johan Engelen, 2010
%% Accompanies image file 'game_graph_2.pdf' (pdf, eps, ps)
%%
%% To include the image in your LaTeX document, write
%%   \input{<filename>.pdf_tex}
%%  instead of
%%   \includegraphics{<filename>.pdf}
%% To scale the image, write
%%   \def\svgwidth{<desired width>}
%%   \input{<filename>.pdf_tex}
%%  instead of
%%   \includegraphics[width=<desired width>]{<filename>.pdf}
%%
%% Images with a different path to the parent latex file can
%% be accessed with the `import' package (which may need to be
%% installed) using
%%   \usepackage{import}
%% in the preamble, and then including the image with
%%   \import{<path to file>}{<filename>.pdf_tex}
%% Alternatively, one can specify
%%   \graphicspath{{<path to file>/}}
%% 
%% For more information, please see info/svg-inkscape on CTAN:
%%   http://tug.ctan.org/tex-archive/info/svg-inkscape
%%
\begingroup%
  \makeatletter%
  \providecommand\color[2][]{%
    \errmessage{(Inkscape) Color is used for the text in Inkscape, but the package 'color.sty' is not loaded}%
    \renewcommand\color[2][]{}%
  }%
  \providecommand\transparent[1]{%
    \errmessage{(Inkscape) Transparency is used (non-zero) for the text in Inkscape, but the package 'transparent.sty' is not loaded}%
    \renewcommand\transparent[1]{}%
  }%
  \providecommand\rotatebox[2]{#2}%
  \newcommand*\fsize{\dimexpr\f@size pt\relax}%
  \newcommand*\lineheight[1]{\fontsize{\fsize}{#1\fsize}\selectfont}%
  \ifx\svgwidth\undefined%
    \setlength{\unitlength}{690bp}%
    \ifx\svgscale\undefined%
      \relax%
    \else%
      \setlength{\unitlength}{\unitlength * \real{\svgscale}}%
    \fi%
  \else%
    \setlength{\unitlength}{\svgwidth}%
  \fi%
  \global\let\svgwidth\undefined%
  \global\let\svgscale\undefined%
  \makeatother%
  \begin{picture}(1,0.52173913)%
    \lineheight{1}%
    \setlength\tabcolsep{0pt}%
    \put(0,0){\includegraphics[width=\unitlength,page=1]{game_graph_2.pdf}}%
    \put(0.71784728,0.03550056){\color[rgb]{0,0,0}\makebox(0,0)[t]{\lineheight{1.25}\smash{\begin{tabular}[t]{c}$T_l+T_l$\end{tabular}}}}%
    \put(0.89199366,0.03550056){\color[rgb]{0,0,0}\makebox(0,0)[t]{\lineheight{1.25}\smash{\begin{tabular}[t]{c}$T_l+T$\end{tabular}}}}%
    \put(0,0){\includegraphics[width=\unitlength,page=2]{game_graph_2.pdf}}%
    \put(0.05764442,0.03425097){\color[rgb]{0,0,0}\makebox(0,0)[t]{\lineheight{1.25}\smash{\begin{tabular}[t]{c}$0$\end{tabular}}}}%
    \put(0.38595504,0.03550056){\color[rgb]{0,0,0}\makebox(0,0)[t]{\lineheight{1.25}\smash{\begin{tabular}[t]{c}$T_l$\end{tabular}}}}%
    \put(0.55011037,0.03425097){\color[rgb]{0,0,0}\makebox(0,0)[t]{\lineheight{1.25}\smash{\begin{tabular}[t]{c}$T$\end{tabular}}}}%
    \put(0.22448655,0.03423782){\color[rgb]{0,0,0}\makebox(0,0)[t]{\lineheight{1.25}\smash{\begin{tabular}[t]{c}$\dots$\end{tabular}}}}%
    \put(0.55104364,0.11638731){\color[rgb]{0,0,0}\makebox(0,0)[rt]{\lineheight{1.25}\smash{\begin{tabular}[t]{r}$\mathbf{z}^{i, *}_{T_l:T_l+T_l-1}$\end{tabular}}}}%
    \put(0,0){\includegraphics[width=\unitlength,page=3]{game_graph_2.pdf}}%
    \put(0.84066617,0.50197625){\color[rgb]{0,0,0}\makebox(0,0)[rt]{\lineheight{1.25}\smash{\begin{tabular}[t]{r}$\mathbf{z}^{i, *}_{T_l:T_l+T-1}$\end{tabular}}}}%
    \put(0,0){\includegraphics[width=\unitlength,page=4]{game_graph_2.pdf}}%
  \end{picture}%
\endgroup%

%% file: figures/time-padding.pdf_tex
%% Creator: Inkscape 1.2.2 (b0a8486541, 2022-12-01), www.inkscape.org
%% PDF/EPS/PS + LaTeX output extension by Johan Engelen, 2010
%% Accompanies image file 'time-padding.pdf' (pdf, eps, ps)
%%
%% To include the image in your LaTeX document, write
%%   \input{<filename>.pdf_tex}
%%  instead of
%%   \includegraphics{<filename>.pdf}
%% To scale the image, write
%%   \def\svgwidth{<desired width>}
%%   \input{<filename>.pdf_tex}
%%  instead of
%%   \includegraphics[width=<desired width>]{<filename>.pdf}
%%
%% Images with a different path to the parent latex file can
%% be accessed with the `import' package (which may need to be
%% installed) using
%%   \usepackage{import}
%% in the preamble, and then including the image with
%%   \import{<path to file>}{<filename>.pdf_tex}
%% Alternatively, one can specify
%%   \graphicspath{{<path to file>/}}
%% 
%% For more information, please see info/svg-inkscape on CTAN:
%%   http://tug.ctan.org/tex-archive/info/svg-inkscape
%%
\begingroup%
  \makeatletter%
  \providecommand\color[2][]{%
    \errmessage{(Inkscape) Color is used for the text in Inkscape, but the package 'color.sty' is not loaded}%
    \renewcommand\color[2][]{}%
  }%
  \providecommand\transparent[1]{%
    \errmessage{(Inkscape) Transparency is used (non-zero) for the text in Inkscape, but the package 'transparent.sty' is not loaded}%
    \renewcommand\transparent[1]{}%
  }%
  \providecommand\rotatebox[2]{#2}%
  \newcommand*\fsize{\dimexpr\f@size pt\relax}%
  \newcommand*\lineheight[1]{\fontsize{\fsize}{#1\fsize}\selectfont}%
  \ifx\svgwidth\undefined%
    \setlength{\unitlength}{375.00002289bp}%
    \ifx\svgscale\undefined%
      \relax%
    \else%
      \setlength{\unitlength}{\unitlength * \real{\svgscale}}%
    \fi%
  \else%
    \setlength{\unitlength}{\svgwidth}%
  \fi%
  \global\let\svgwidth\undefined%
  \global\let\svgscale\undefined%
  \makeatother%
  \begin{picture}(1,0.49999997)%
    \lineheight{1}%
    \setlength\tabcolsep{0pt}%
    \put(0,0){\includegraphics[width=\unitlength,page=1]{time-padding.pdf}}%
    \put(0.19374793,0.21050681){\color[rgb]{0,0,0}\makebox(0,0)[rt]{\lineheight{1.25}\smash{\begin{tabular}[t]{r}\tiny$\mathbf{z}^{i*}_{t:t+T-1}$\end{tabular}}}}%
    \put(0.28710114,0.1426165){\color[rgb]{0.17254902,0.63529412,0.37254902}\makebox(0,0)[t]{\lineheight{1.25}\smash{\begin{tabular}[t]{c}\textsf{\footnotesize{executed}}\end{tabular}}}}%
    \put(0.89949728,0.16439016){\color[rgb]{0.70196078,0.70196078,0.70196078}\makebox(0,0)[t]{\lineheight{1.25}\smash{\begin{tabular}[t]{c}\textsf{\footnotesize{predicted}}\end{tabular}}}}%
    \put(0.85911916,0.24002949){\color[rgb]{0.70196078,0.70196078,0.70196078}\makebox(0,0)[lt]{\lineheight{1.25}\smash{\begin{tabular}[t]{l}\tiny$z^i_{t+T-1}$\end{tabular}}}}%
    \put(0.71664719,0.34123162){\color[rgb]{0.70196078,0.70196078,0.70196078}\makebox(0,0)[t]{\lineheight{1.25}\smash{\begin{tabular}[t]{c}\textsf{\footnotesize{predicted}}\end{tabular}}}}%
    \put(0.19430714,0.28116908){\color[rgb]{0,0,0}\makebox(0,0)[rt]{\lineheight{1.25}\smash{\begin{tabular}[t]{r}\tiny$\mathbf{z}^{i}_{t:t+T-1}$\end{tabular}}}}%
    \put(0.37654021,0.09675207){\color[rgb]{0.50196078,0.50196078,0.50196078}\makebox(0,0)[rt]{\lineheight{1.25}\smash{\begin{tabular}[t]{r}\tiny$\mathbf{z}^i_{t+T_l:t+T+T_l-1}$\end{tabular}}}}%
    \put(0.10425109,0.32653549){\color[rgb]{0.52941176,0.87058824,0.67843137}\makebox(0,0)[t]{\lineheight{1.25}\smash{\begin{tabular}[t]{c}\textsf{\footnotesize{executed}}\end{tabular}}}}%
    \put(0.66631886,0.41534931){\color[rgb]{0.70196078,0.70196078,0.70196078}\makebox(0,0)[lt]{\lineheight{1.25}\smash{\begin{tabular}[t]{l}\tiny$z^i_{t+T-T_l-1}$\end{tabular}}}}%
    \put(0.37654022,0.03275201){\color[rgb]{0.50196078,0.50196078,0.50196078}\makebox(0,0)[rt]{\lineheight{1.25}\smash{\begin{tabular}[t]{r}\tiny$\mathbf{z}^{i*}_{t+T_l:t+T+T_l-1}$\end{tabular}}}}%
    \put(0,0){\includegraphics[width=\unitlength,page=2]{time-padding.pdf}}%
    \put(0.58927892,0.21357672){\color[rgb]{0,0,0}\makebox(0,0)[t]{\lineheight{1.25}\smash{\begin{tabular}[t]{c}\tiny$T-T_l$\end{tabular}}}}%
    \put(0.28510757,0.21357672){\color[rgb]{0,0,0}\makebox(0,0)[t]{\lineheight{1.25}\smash{\begin{tabular}[t]{c}\tiny$T_l$\end{tabular}}}}%
    \put(0.71371705,0.28215004){\color[rgb]{0,0,0}\makebox(0,0)[t]{\lineheight{1.25}\smash{\begin{tabular}[t]{c}\tiny$T_l$\end{tabular}}}}%
    \put(0.40980362,0.28215004){\color[rgb]{0,0,0}\makebox(0,0)[t]{\lineheight{1.25}\smash{\begin{tabular}[t]{c}\tiny$T-T_l$\end{tabular}}}}%
    \put(0.49792332,0.24591716){\color[rgb]{0,0,0}\makebox(0,0)[t]{\lineheight{1.25}\smash{\begin{tabular}[t]{c}\scriptsize{\textsf{IBR}}\end{tabular}}}}%
    \put(0,0){\includegraphics[width=\unitlength,page=3]{time-padding.pdf}}%
    \put(0.77364401,0.02907787){\color[rgb]{0.50196078,0.50196078,0.50196078}\makebox(0,0)[t]{\lineheight{1.25}\smash{\begin{tabular}[t]{c}\tiny$T-T_l$\end{tabular}}}}%
    \put(0.46947266,0.02907787){\color[rgb]{0.50196078,0.50196078,0.50196078}\makebox(0,0)[t]{\lineheight{1.25}\smash{\begin{tabular}[t]{c}\tiny$T_l$\end{tabular}}}}%
    \put(0.89808214,0.09765118){\color[rgb]{0.50196078,0.50196078,0.50196078}\makebox(0,0)[t]{\lineheight{1.25}\smash{\begin{tabular}[t]{c}\tiny$T_l$\end{tabular}}}}%
    \put(0.59416872,0.09765118){\color[rgb]{0.50196078,0.50196078,0.50196078}\makebox(0,0)[t]{\lineheight{1.25}\smash{\begin{tabular}[t]{c}\tiny$T-T_l$\end{tabular}}}}%
    \put(0.68213328,0.06181669){\color[rgb]{0.50196078,0.50196078,0.50196078}\makebox(0,0)[t]{\lineheight{1.25}\smash{\begin{tabular}[t]{c}\scriptsize{\textsf{IBR}}\end{tabular}}}}%
    \put(0,0){\includegraphics[width=\unitlength,page=4]{time-padding.pdf}}%
    \put(0.40423296,0.39557872){\color[rgb]{0.50196078,0.50196078,0.50196078}\makebox(0,0)[t]{\lineheight{1.25}\smash{\begin{tabular}[t]{c}\tiny$T-T_l$\end{tabular}}}}%
    \put(0.10006161,0.39557872){\color[rgb]{0.50196078,0.50196078,0.50196078}\makebox(0,0)[t]{\lineheight{1.25}\smash{\begin{tabular}[t]{c}\tiny$T_l$\end{tabular}}}}%
    \put(0.52867109,0.46415203){\color[rgb]{0.50196078,0.50196078,0.50196078}\makebox(0,0)[t]{\lineheight{1.25}\smash{\begin{tabular}[t]{c}\tiny$T_l$\end{tabular}}}}%
    \put(0.22475766,0.46415203){\color[rgb]{0.50196078,0.50196078,0.50196078}\makebox(0,0)[t]{\lineheight{1.25}\smash{\begin{tabular}[t]{c}\tiny$T-T_l$\end{tabular}}}}%
    \put(0.31285233,0.42807406){\color[rgb]{0.50196078,0.50196078,0.50196078}\makebox(0,0)[t]{\lineheight{1.25}\smash{\begin{tabular}[t]{c}\scriptsize{\textsf{IBR}}\end{tabular}}}}%
    \put(0,0){\includegraphics[width=\unitlength,page=5]{time-padding.pdf}}%
  \end{picture}%
\endgroup%

%% file: figures/field_scenario.pdf_tex
%% Creator: Inkscape 1.4.2 (ebf0e940d0, 2025-05-08), www.inkscape.org
%% PDF/EPS/PS + LaTeX output extension by Johan Engelen, 2010
%% Accompanies image file 'field_scenario.pdf' (pdf, eps, ps)
%%
%% To include the image in your LaTeX document, write
%%   \input{<filename>.pdf_tex}
%%  instead of
%%   \includegraphics{<filename>.pdf}
%% To scale the image, write
%%   \def\svgwidth{<desired width>}
%%   \input{<filename>.pdf_tex}
%%  instead of
%%   \includegraphics[width=<desired width>]{<filename>.pdf}
%%
%% Images with a different path to the parent latex file can
%% be accessed with the `import' package (which may need to be
%% installed) using
%%   \usepackage{import}
%% in the preamble, and then including the image with
%%   \import{<path to file>}{<filename>.pdf_tex}
%% Alternatively, one can specify
%%   \graphicspath{{<path to file>/}}
%% 
%% For more information, please see info/svg-inkscape on CTAN:
%%   http://tug.ctan.org/tex-archive/info/svg-inkscape
%%
\begingroup%
  \makeatletter%
  \providecommand\color[2][]{%
    \errmessage{(Inkscape) Color is used for the text in Inkscape, but the package 'color.sty' is not loaded}%
    \renewcommand\color[2][]{}%
  }%
  \providecommand\transparent[1]{%
    \errmessage{(Inkscape) Transparency is used (non-zero) for the text in Inkscape, but the package 'transparent.sty' is not loaded}%
    \renewcommand\transparent[1]{}%
  }%
  \providecommand\rotatebox[2]{#2}%
  \newcommand*\fsize{\dimexpr\f@size pt\relax}%
  \newcommand*\lineheight[1]{\fontsize{\fsize}{#1\fsize}\selectfont}%
  \ifx\svgwidth\undefined%
    \setlength{\unitlength}{513.77952756bp}%
    \ifx\svgscale\undefined%
      \relax%
    \else%
      \setlength{\unitlength}{\unitlength * \real{\svgscale}}%
    \fi%
  \else%
    \setlength{\unitlength}{\svgwidth}%
  \fi%
  \global\let\svgwidth\undefined%
  \global\let\svgscale\undefined%
  \makeatother%
  \begin{picture}(1,0.60689655)%
    \lineheight{1}%
    \setlength\tabcolsep{0pt}%
    \put(0,0){\includegraphics[width=\unitlength,page=1]{field_scenario.pdf}}%
    \put(0.86322029,0.52577991){\color[rgb]{0,0,0}\transparent{0.80000001}\makebox(0,0)[t]{\lineheight{1.25}\smash{\begin{tabular}[t]{c}\small{$\succcurlyeq_J$}\\\\\end{tabular}}}}%
    \put(0.85759151,0.41839046){\color[rgb]{0,0,0}\transparent{0.80000001}\makebox(0,0)[t]{\lineheight{1.25}\smash{\begin{tabular}[t]{c}\textsf{\small reach goal}\\\end{tabular}}}}%
    \put(0.85759151,0.48800788){\color[rgb]{0,0,0}\transparent{0.80000001}\makebox(0,0)[t]{\lineheight{1.25}\smash{\begin{tabular}[t]{c}\textsf{\small stay in lane}\\\end{tabular}}}}%
    \put(0.85759151,0.55618888){\color[rgb]{0,0,0}\transparent{0.80000001}\makebox(0,0)[t]{\lineheight{1.25}\smash{\begin{tabular}[t]{c}\textsf{\small not brake}\\\end{tabular}}}}%
    \put(0.86322029,0.45616251){\color[rgb]{0,0,0}\transparent{0.80000001}\makebox(0,0)[t]{\lineheight{1.25}\smash{\begin{tabular}[t]{c}\small{$\succcurlyeq_J$}\\\\\end{tabular}}}}%
    \put(0.14915888,0.4528101){\color[rgb]{0,0,0}\transparent{0.80000001}\makebox(0,0)[t]{\lineheight{1.25}\smash{\begin{tabular}[t]{c}\small{$\succcurlyeq_J$}\\\\\end{tabular}}}}%
    \put(0.1442307,0.41400959){\color[rgb]{0,0,0}\transparent{0.80000001}\makebox(0,0)[t]{\lineheight{1.25}\smash{\begin{tabular}[t]{c}\textsf{\small stay on road}\\\end{tabular}}}}%
    \put(0.14423069,0.49152302){\color[rgb]{0,0,0}\transparent{0.80000001}\makebox(0,0)[t]{\lineheight{1.25}\smash{\begin{tabular}[t]{c}\textsf{\small reach goal}\\\end{tabular}}}}%
    \put(0.16731957,0.13509378){\color[rgb]{0,0,0}\transparent{0.80000001}\makebox(0,0)[t]{\lineheight{1.25}\smash{\begin{tabular}[t]{c}\small{$\succcurlyeq_J$}\\\\\end{tabular}}}}%
    \put(0.1616907,0.02770432){\color[rgb]{0,0,0}\transparent{0.80000001}\makebox(0,0)[t]{\lineheight{1.25}\smash{\begin{tabular}[t]{c}\textsf{\small reach goal}\\\end{tabular}}}}%
    \put(0.16169071,0.09732174){\color[rgb]{0,0,0}\transparent{0.80000001}\makebox(0,0)[t]{\lineheight{1.25}\smash{\begin{tabular}[t]{c}\textsf{\small stay in lane}\\\end{tabular}}}}%
    \put(0.16169071,0.16550276){\color[rgb]{0,0,0}\transparent{0.80000001}\makebox(0,0)[t]{\lineheight{1.25}\smash{\begin{tabular}[t]{c}\textsf{\small not brake}\\\end{tabular}}}}%
    \put(0.16731957,0.06547637){\color[rgb]{0,0,0}\transparent{0.80000001}\makebox(0,0)[t]{\lineheight{1.25}\smash{\begin{tabular}[t]{c}\small{$\succcurlyeq_J$}\\\end{tabular}}}}%
    \put(0.63706724,0.06490192){\color[rgb]{0,0,0}\transparent{0.80000001}\makebox(0,0)[lt]{\lineheight{1.25}\smash{\begin{tabular}[t]{l}\footnotesize\textsf{prediction}\end{tabular}}}}%
    \put(0.43011302,0.08865651){\color[rgb]{0,0,0}\transparent{0.80000001}\makebox(0,0)[lt]{\lineheight{1.25}\smash{\begin{tabular}[t]{l}\footnotesize\textsf{previous}\end{tabular}}}}%
    \put(0,0){\includegraphics[width=\unitlength,page=2]{field_scenario.pdf}}%
    \put(0.43011302,0.05946108){\color[rgb]{0,0,0}\transparent{0.80000001}\makebox(0,0)[lt]{\lineheight{1.25}\smash{\begin{tabular}[t]{l}\footnotesize\textsf{solution}\end{tabular}}}}%
    \put(0,0){\includegraphics[width=\unitlength,page=3]{field_scenario.pdf}}%
  \end{picture}%
\endgroup%

%% file: figures/city_scenario.pdf_tex
%% Creator: Inkscape 1.4.2 (ebf0e940d0, 2025-05-08), www.inkscape.org
%% PDF/EPS/PS + LaTeX output extension by Johan Engelen, 2010
%% Accompanies image file 'city_scenario.pdf' (pdf, eps, ps)
%%
%% To include the image in your LaTeX document, write
%%   \input{<filename>.pdf_tex}
%%  instead of
%%   \includegraphics{<filename>.pdf}
%% To scale the image, write
%%   \def\svgwidth{<desired width>}
%%   \input{<filename>.pdf_tex}
%%  instead of
%%   \includegraphics[width=<desired width>]{<filename>.pdf}
%%
%% Images with a different path to the parent latex file can
%% be accessed with the `import' package (which may need to be
%% installed) using
%%   \usepackage{import}
%% in the preamble, and then including the image with
%%   \import{<path to file>}{<filename>.pdf_tex}
%% Alternatively, one can specify
%%   \graphicspath{{<path to file>/}}
%% 
%% For more information, please see info/svg-inkscape on CTAN:
%%   http://tug.ctan.org/tex-archive/info/svg-inkscape
%%
\begingroup%
  \makeatletter%
  \providecommand\color[2][]{%
    \errmessage{(Inkscape) Color is used for the text in Inkscape, but the package 'color.sty' is not loaded}%
    \renewcommand\color[2][]{}%
  }%
  \providecommand\transparent[1]{%
    \errmessage{(Inkscape) Transparency is used (non-zero) for the text in Inkscape, but the package 'transparent.sty' is not loaded}%
    \renewcommand\transparent[1]{}%
  }%
  \providecommand\rotatebox[2]{#2}%
  \newcommand*\fsize{\dimexpr\f@size pt\relax}%
  \newcommand*\lineheight[1]{\fontsize{\fsize}{#1\fsize}\selectfont}%
  \ifx\svgwidth\undefined%
    \setlength{\unitlength}{513.77952756bp}%
    \ifx\svgscale\undefined%
      \relax%
    \else%
      \setlength{\unitlength}{\unitlength * \real{\svgscale}}%
    \fi%
  \else%
    \setlength{\unitlength}{\svgwidth}%
  \fi%
  \global\let\svgwidth\undefined%
  \global\let\svgscale\undefined%
  \makeatother%
  \begin{picture}(1,0.60689655)%
    \lineheight{1}%
    \setlength\tabcolsep{0pt}%
    \put(0,0){\includegraphics[width=\unitlength,page=1]{city_scenario.pdf}}%
    \put(0.16399882,0.13447021){\color[rgb]{0,0,0}\transparent{0.80000001}\makebox(0,0)[t]{\lineheight{1.25}\smash{\begin{tabular}[t]{c}\small{$\succcurlyeq_J$}\\\\\end{tabular}}}}%
    \put(0.16128947,0.03161558){\color[rgb]{0,0,0}\transparent{0.80000001}\makebox(0,0)[t]{\lineheight{1.25}\smash{\begin{tabular}[t]{c}\textsf{\small reach goal}\\\end{tabular}}}}%
    \put(0.16128947,0.16649452){\color[rgb]{0,0,0}\transparent{0.80000001}\makebox(0,0)[t]{\lineheight{1.25}\smash{\begin{tabular}[t]{c}\textsf{\small stay in lane}\\\end{tabular}}}}%
    \put(0.16128948,0.09583854){\color[rgb]{0,0,0}\transparent{0.80000001}\makebox(0,0)[t]{\lineheight{1.25}\smash{\begin{tabular}[t]{c}\textsf{\small not brake}\\\end{tabular}}}}%
    \put(0.87510447,0.52605289){\color[rgb]{0,0,0}\transparent{0.80000001}\makebox(0,0)[t]{\lineheight{1.25}\smash{\begin{tabular}[t]{c}\small{$\succcurlyeq_J$}\\\\\end{tabular}}}}%
    \put(0.87239519,0.48742828){\color[rgb]{0,0,0}\transparent{0.80000001}\makebox(0,0)[t]{\lineheight{1.25}\smash{\begin{tabular}[t]{c}\textsf{\small not brake}\\\end{tabular}}}}%
    \put(0.87239519,0.4215049){\color[rgb]{0,0,0}\transparent{0.80000001}\makebox(0,0)[t]{\lineheight{1.25}\smash{\begin{tabular}[t]{c}\textsf{\small reach goal}\\\end{tabular}}}}%
    \put(0.87239519,0.55807721){\color[rgb]{0,0,0}\transparent{0.80000001}\makebox(0,0)[t]{\lineheight{1.25}\smash{\begin{tabular}[t]{c}\textsf{\small stay in lane}\\\end{tabular}}}}%
    \put(0.87510447,0.45832339){\color[rgb]{0,0,0}\transparent{0.80000001}\makebox(0,0)[t]{\lineheight{1.25}\smash{\begin{tabular}[t]{c}\small{$\succcurlyeq_J$}\\\\\end{tabular}}}}%
    \put(0.14538676,0.49632948){\color[rgb]{0,0,0}\transparent{0.80000001}\makebox(0,0)[t]{\lineheight{1.25}\smash{\begin{tabular}[t]{c}\textsf{\small stay on road}\\\end{tabular}}}}%
    \put(0.14538675,0.42456701){\color[rgb]{0,0,0}\transparent{0.80000001}\makebox(0,0)[t]{\lineheight{1.25}\smash{\begin{tabular}[t]{c}\textsf{\small reach goal}\\\end{tabular}}}}%
    \put(0.14855155,0.46341158){\color[rgb]{0,0,0}\transparent{0.80000001}\makebox(0,0)[t]{\lineheight{1.25}\smash{\begin{tabular}[t]{c}\small{$\succcurlyeq_J$}\\\\\end{tabular}}}}%
    \put(0.16474287,0.06628986){\color[rgb]{0,0,0}\transparent{0.80000001}\makebox(0,0)[t]{\lineheight{1.25}\smash{\begin{tabular}[t]{c}\small{$\succcurlyeq_J$}\\\\\end{tabular}}}}%
    \put(0,0){\includegraphics[width=\unitlength,page=2]{city_scenario.pdf}}%
  \end{picture}%
\endgroup%